\documentstyle[12pt]{article}

\begin{document}
\newcommand {\be}{\begin{equation}}
\newcommand {\ee}{\end{equation}}
\newcommand {\bea}{\begin{array}}
\newcommand {\cl}{\centerline}
\newcommand {\eea}{\end{array}}
\renewcommand {\theequation}{\thesection.\arabic{equation}}
\newcommand {\newsection}{\setcounter{equation}{0}\section}
\def\nct{noncommutative torus }
\def\ncy{noncommutativity }
\def\comp{compactification }
\def\nc{noncommutative }
\def\quan{quantization }
\def\BC{boundary condition }
\def\BCs{boundary conditions }
\def\cons{constraints }
\baselineskip 0.65 cm
\begin{flushright}
IPM/P-99/035 \\
hep-th/9906161
\end{flushright}
\begin{center}
 {\Large {\bf  Dirac Quantization of Open Strings and Noncommutativity in Branes}}
\vskip .5cm

F. Ardalan$^{a,b}$, H. Arfaei$^{a,b}$ and M.M. Sheikh-Jabbari$^a$
\footnote{ E-mail:  ardalan, arfaei, jabbari@theory.ipm.ac.ir } \\

\vskip .5cm

 {\it $^a$ Institute for Studies in Theoretical Physics and Mathematics 
IPM,

 P.O.Box 19395-5531, Tehran, Iran}\\
{\it and}
\\
{\it $^b$ Department of Physics Sharif University of Technology, 

P.O.Box 11365-9161, Tehran, Iran}
\end{center}

\vskip 2cm
\begin{abstract}
We apply the Dirac bracket quantization to open strings attached to branes 
in the presence of background antisymmetric field and recover an inherent 
\ncy in the internal coordinates of the brane.

\end{abstract}
\newpage

\section{Introduction}
\setcounter{equation}{0} 
   Since the discovery of the role of branes in string theory [1] they have 
frequently shown unexpected properties. They were first identified as
the carriers of R-R charges and very soon after, it was realized that when 
$N$ of them merge the space-time coordinates normal to them become noncommutative
[2] and the $U(N)$ Super-Yang-Mills theory emerges. 
Another type of \ncy appears in the bound states of branes with fundamental 
strings and with other branes
\footnote{ Another type of \ncy appeared in the context f 2+1 dimensional
Chern-Simons theory [26].}. It has been shown that such brane bound states 
correspond to branes
with non-zero background internal gauge fields [2,3,4]. 
These are the type of branes we shall be focusing on in this article.
The correspondence has been demonstrated 
by constructing the interaction amplitudes of 
two branes [4,5] and by looking at the scattering of different closed
strings off the branes [6]. 
The \ncy arising in the internal structure of these brane bound states 
is  a consequence of  the properties of open strings ending on them. 
Such open strings satisfy boundary conditions which are neither Neumann nor 
Dirichlet, but a combination of the two, sometimes referred to as mixed boundary 
condition [4,5,7].
The mixed boundary condition makes the canonical quantization of the theory 
non-trivial. Imposing the standard commutation relations, leads to
inconsistency. It has been proposed to remove the inconsistency by relaxing 
the commutativity of the space coordinates of the open strings along the 
direction of the brane described by mixed boundary conditions [7,8,9].
The procedure of relaxing the commutativity of space coordinates adopted in 
[5,8], was to keep the standard
algebra of the Fourier modes in the mode expansion. 

The noncommutativity observed in the above brane system seemed  very
similar to that observed by Connes, Douglas, Schwartz (CDS) [10] in the  
problem of Matrix Model with non-trivial background three form. 
They studied compactification of Matrix theory on a noncommutative torus and
realized that it corresponds to the Matrix theory in such backgrounds. 

Motivated by this observation, we showed that [7,8],
the \ncy can be derived within the string
theory by wrapping branes with non-zero $B_{ij}$ background field on the
\comp torus.
Moreover in [8]  the $C^*$-algebra of functions on
noncommutative torus which is building block of the Connes' noncommutative 
geometry construction, [10,11], was also constructed 
using the noncommutative open string position operators.  

The BPS spectrum of Yang-Mills theory on noncommutative torus was also 
originally obtained by CDS. Later the spectrum of the \nc M(atrix)-model
was derived by taking the longitudinal and transverse membrane winding into 
account [12]. 
As another evidence for the above correspondence, in [8,13] it was shown that
the low energy BPS spectrum of such brane system is exactly the same as 
the results of [10,12].

Soon after [10], another check on this correspondence was made by Douglas and 
Hull [14]. They showed that string compactification, along with 
wrapping a two-brane in the non-zero two-from background, $B_{ij}$, in the low
energy limit leads to Yang-Mills theory on the dual \nc torus. 
Studying the three and four open strings scattering amplitudes in [15],  
the noncommutative Yang-Mills theory  was obtained as 
the related low energy effective field theory on the branes.
Further evidence for the correspondence of the work of CDS and string theory 
in a non-zero $B_{\mu \nu}$ background field has been discussed in many 
papers [16].
Despite these checks which support the noncommutativity of the different
position coordinates of the brane with non zero background B-field, the
arguments leading to noncommutativity of space coordinates was based on 
a number of assumptions which might be open to question.
The method adopted in [5,8] has been based on taking the commutation relation 
among the raising and lowering operators of the open strings in absence of the
$B_{\mu\nu}$ field and imposing the commutativity on the zero modes. 
On the other hand in [9] the analysis was based a time
averaged symplectic form. 

In this article we will try to provide a solid ground for the quantization
of this system. Since the serious obstacle in the way of quantization is
the mixed boundary condition we need to treat it carefully. We consider it
as a constraint and apply the method of Dirac to find the commutators. 
In the rest of the introduction we give  a brief review of the related
matters, the Matrix model on the noncommutative torus, the previous
approaches to the quantization of strings attached to branes with
background B fields, and an outline of the Dirac method [17].  

In the M(atrix)-model one considers a set of $N\times N$ bosonic hermitian 
matrices, $X^a,\; a=1,...,9$ and sixteen component spinor of $SO(9)$,
$\Psi$, and dynamics of these matrices is given by
\be
I={1 \over 2g\sqrt{\alpha'}}\int d\tau \;\;\;Tr \; \biggl\{ \dot{X_a}\dot{X_a}
+{1 \over {(2\pi\alpha')}^2}\sum_{a<b}[X^a,X^b]^2 \\
+{i \over 2\pi\alpha'}\Psi^T\dot{\Psi}-{1 \over (2\pi\alpha')^2}\Psi^T
\Gamma_a[X^a,\Psi]\biggr\}.
\ee
M(atrix) model was conjectured to be a light cone formulation of M-theory [18]. 
Thus, although the theory originally involves eleven dimensions, it has
nine dimensional objects as functions of time. 
The above conjecture has passed many consistency checks [19], one of the most
important of them was its compactifications on different manifolds.
Compactification of the matrix coordinate $X^a$ on a circle of radius $R$ is 
possible if there exist matrices $U$ as a part of the symmetries of (1.1)
that shifts $X^a$ by $R$;
\be
UX^aU^{-1}=X^a+R
\ee
leaving the other coordinates and spinors unchanged. Modding out the action of $U$
is equivalent to \comp. This $S^1$ compactification leads to the
string Matrix theory of Dijkgraaf-Verlinde-Verlinde [20] which is shown to
be 
equivalent to type IIA free string theory.
Compactification on higher dimensional tori is given similarly;
\be
\bea{cc}
U_iX_iU_i^{-1}=X_i+R_i, \\
U_iX_j U_i^{-1}=X_j  \;\;\;\;\;\; i\neq j ,\\
U_i\Psi U_i^{-1}=\Psi.
\eea
\ee
Consistency requires that $U_i$'s should commute up to  a phase;
\be
U_iU_j=e^{i\theta_{ij}}U_jU_i.
\ee
For ordinary commutative torus $U_i$'s are translation operators with 
$\theta_{ij}=0$
\footnote{Due to SL(2,Z)-type symmetries [11] rational $\theta_{ij}$'s can be  
mapped to an integer and hence they also result in commutative space.}.
If $\sigma_i$'s parameterize the dual torus, a solution to (1.3) for 
$\theta_{ij}=0$ is
\be \bea{cc}
X_i=i{\partial \over \partial \sigma_i}+ A_i(\sigma), \\
U_i=e^{i\sigma_i R_i},
\eea \ee
It is easily seen that such \comp yields the Yang-Mills theory on the dual torus. 

The major step taken by CDS, [10], was to realize that non-zero, irrational   
$\theta_{ij}$ leads to \comp on a \nct, and the solution to (1.3) at the
limit
of large $N$ describes a Yang-Mills theory on the dual \nct in which we have 
to replace the definition of commutators with the Moyal brackets:
\be\bea{cc}
\bigl\{ A,B\bigr\}=A*B-B*A ,\\
A*B(\sigma)=e^{-i\theta(\partial'_1\partial''_2-\partial'_2\partial''_1)} A(\sigma')B(\sigma'')|_
{\sigma'=\sigma''=\sigma}.
\eea \ee
Non-zero $\theta_{ij}$ leads to non-locality in the gauge theory and break down 
of Lorentz invariance. This has led to a line of investigation on Moyal gauge 
theories [21]. In their original work CDS attempt to identify physical 
sources of \ncy and suggest that it is due to non-zero $C_{-ij}$,
the three form as a background, when one formulates the M-theory on the light 
cone.

We will now review the two previous approaches
of mixed open strings quantization. We take the open strings
having their ends on non-marginal BPS bound states of p, p-2, p-4,... branes. 
Such bound states are represented by a non-zero static and 
uniform configuration of $B_{\mu\nu}$ field in the bulk and $A_{\mu}$
field with constant strength, $F_{\mu\nu}$, in the brane. 
Suppose the brane spans a
p+1 dimensional space $(X^0,X^1,....,X^p)$ with $X^{p+1}=X^{p+2}=...=X^9=0$.
Action for the open string with its ends on this brane is [22];

\be\bea{cc}
S= {1 \over 4\pi\alpha'} \int_{\Sigma} d^2\sigma \bigl[ \eta_{\mu\nu}
\partial_aX^{\mu}\partial_bX^{\nu}g^{ab}+ \epsilon^{ab} B_{\mu\nu}\partial_a
X^{\mu}\partial_bX^{\nu}\bigr]+\\ 
{1 \over 2\pi\alpha'}\int d \tau A_i \partial_{\tau}X^i|_{\pi}- 
{1 \over 2\pi\alpha'}\int d \tau A_i \partial_{\tau}X^i|_{0}
\eea\ee
where $A_i,\ i=0,1,...,p$ is the $U(1)$ gauge field living on the D-brane.
The symmetries of the action includes on top of the geometric symmetries,
two $U(1)$ gauge symmetries, one of them, which we call it $\lambda$-gauge 
symmetry, only acts on the internal field:
\be
A\rightarrow A+d\lambda,
\ee
and this is the symmetry which  raises to $U(N)$ symmetry when $N$ branes
coincide. The other $U(1)$ symmetry, which we will note it by $\Lambda$-gauge 
symmetry, acts on both $B_{\mu\nu}$ and $A_i$:
\be \bea{cc}
B_{\mu\nu}\rightarrow B_{\mu\nu}+\partial_{\mu}\Lambda_{\nu}-\partial_{\nu}\Lambda_{\mu} \\
A_{\mu} \rightarrow A_{\mu}+\Lambda_{\mu},
\eea\ee
and hence the gauge invariant field strength is 
\be
{\cal F}_{\mu\nu}=B_{\mu\nu}-F_{\mu\nu} \;\;\;\;\; ,\;\; 
F_{\mu\nu}=dA=\partial_{[\mu}A_{\nu]}.
\ee

The equations of motion are the same as ordinary open strings but the 
boundary conditions at ends are neither Neumann nor Dirichlet but a linear 
combination of the two,
\be
\partial_{\sigma}X^{\mu}+{\cal F}_{\mu\nu} \partial_{\tau}X^{\nu}=0 \;\;\;
{\rm at}\;\; \sigma=0,\pi.
\ee
To quantize the theory we should first find the canonical conjugates of $X^{\mu}$
\be
P_{\mu}(\sigma,\tau)= {\partial S\over \partial X^{\mu}(\sigma,\tau)}= 
{1\over2\pi\alpha'}\biggl(\partial_{\tau}X_{\mu}+{B}_{\mu\nu} 
\partial_{\sigma}X^{\nu}+A_{\mu}(\delta(\sigma-\pi)-\delta(\sigma)\biggr).
\ee
An important point is the freedom in the definition of the conjugate momentum.
It  depends on the choice of both $U(1)$ gauges discussed above.
Therefore it suggests
that the noncommutativity may be an artifact of these choices. To clarify this  
point, in this paper, we do not fix the gauge and conclude that within Dirac's 
method of quantization although the noncommutativity depends on the gauge, 
it is not removable by a gauge choice. The physical quantities such as  
spectrum depend on the gauge invariant field ${\cal F}$. The gauge choice
of [4,5] and [8] which was also adopted in [9], 
\be
A_{\mu}=0 \;\;\;{\rm or}\;\;\;\  {\cal F}=B,
\ee
has no delta function singularity  at the ends of the string.  
In this gauge the conjugate momenta are
\be
2\pi\alpha'P_{\mu}(\sigma,\tau)=\partial_{\tau}X_{\mu}+{\cal F}_{\mu\nu} 
\partial_{\sigma}X^{\nu}.
\ee
Imposing the canonical quantization conditions proves to be inconsistent and 
hence we have to search for a remedy. 
Although in [8] the concentration was on the compact case, 
the arguments given there hold for the non-compact case too.
The method adopted there is to make the commutation relation of the mixed 
open strings modes satisfy the same  algebra as the Neumann open strings.

Taking 
\be\bea{cc}
[ P^{\mu}(\sigma,\tau),P^{\nu}(\sigma',\tau)]=0, \\   \;\;\;
[X^{\mu}(\sigma,\tau),P^{\nu}(\sigma',\tau)]=
i\eta^{\mu\nu}\delta(\sigma-\sigma'),
\eea\ee
as essential commutation relations, and 
imposing them on  the classical mixed open string mode expansions
\be\bea{cc}
X^{\mu}=x^{\mu}+p^{\mu}\tau-{\cal F}^{\mu\nu}p_{\nu}\sigma+ 
\sum_{n\neq 0} {e^{-in\tau} \over n}
\bigl(ia^{\mu}_n \cos n\sigma -{\cal F}^{\mu}_{\nu}a^{\nu}_n\sin n\sigma \bigr)\\
2\pi\alpha'P^{\mu}={\cal M}^{\mu}_{\nu}\bigl(p^{\mu}+
\sum_{n\neq 0} {e^{-in\tau}}a^{\mu}_n \cos n\sigma\bigr), \;\;\;\
{\cal M}={\bf 1}-{\cal F}^2,
\eea\ee
we find that 
\be
[p^{\mu},p^{\nu}]=0 \;\;\; , \;\;\;
[x^{\mu},p^{\nu}]=2i\alpha'({\cal M}^{-1})^{\mu\nu} \;\;\; ,\;\;\;
[a_n^{\mu},a_m^{\nu}]=2\alpha'({\cal M}^{-1})^{\mu\nu} n\delta_{n+m}.
\ee
Now let us consider  the $X^{\mu}$ , $X^{\nu}$ commutators,
\be
[X^{\mu}(\sigma,\tau),X^{\nu}(\sigma',\tau)]=[x^{\mu},x^{\nu}]+ 
2i\alpha'({\cal M}^{-1}{\cal F})^{\mu\nu}
\bigl(\sigma+\sigma'+ \sum_{n\neq 0}{1\over n}\sin n(\sigma+\sigma')\bigr).
\ee
The sum appearing in the above equation is 
\be
\sigma+\sigma'+ \sum_{n\neq 0}{1\over n}\sin n(\sigma+\sigma')=\left\{\bea{cc}
0 \;\;\;\;\ \sigma=\sigma'=0 \\ 
2\pi \;\;\;\; \sigma=\sigma'=\pi \\
\pi  \;\;\;\;\; otherwise,
\eea\right.
\ee
and hence
\be
[X^{\mu}(\sigma,\tau),X^{\nu}(\sigma',\tau)]=\left\{\bea{cc}
[x^{\mu},x^{\nu}] \;\;\;\;\;\;\;\;\;\;\;\;\;\;\;  \sigma=\sigma'=0 \\ \;
[x^{\mu},x^{\nu}]+ 4\pi i\alpha'({\cal M}^{-1}{\cal F})^{\mu\nu} 
\;\;\;\; \sigma=\sigma'=\pi \\ \;
[x^{\mu},x^{\nu}]+ 2 \pi i\alpha'({\cal M}^{-1}{\cal F})^{\mu\nu} 
\;\;\;\;\; otherwise.
\eea\right.
\ee
In this intuitive quantization method $[x^{\mu},x^{\nu}]$ is not determined and can
be chosen arbitrarily. In [8] $[x^{\mu},x^{\nu}]$ was {\it chosen} to be zero
and hence the open strings center of mass became noncommuting. This choice was
supported by the following physical argument: To find out whether the space is
commuting or not, we should probe it by strings, open or closed. Since the closed 
strings live in the bulk and not only on the branes, they will not show the 
noncommutative structure of the space [6]. In a more intuitive explanation
since the "right" and "left" moving modes of closed strings contribute equally
in the probing processes, the noncommutativity seen by the left movers cancel
those of right movers and so the space probed by closed strings is commutative.
When we use the mixed open strings as probes because of asymmetry in the  
"left" and "right" modes, the space seen by such strings is noncommuting. And
the related physical quantity which shows the space noncommutativity is the
center of mass coordinates.

In another quantization approach, authors of [9] tried to fix the
arbitrariness of
$[x^{\mu},x^{\nu}]$. They proposed to use the "time averaged" symplectic form:
\be
{\tilde \Omega}={1\over 2\alpha'}\{ {\cal M}_{\mu\nu}dp^{\mu}(dx^{\nu}+
{\pi\over 2}{\cal F}_{\nu\alpha}dp^{\alpha})\}+
\sum_{n>0}{-i\over n}({\cal M}_{\mu\nu}da_n^{\mu}d_n^{\nu}).
\ee
This method leads to 
\be
[x^{\mu},x^{\nu}]=-2\pi i\alpha'({\cal M}^{-1}{\cal F})^{\mu\nu}.
\ee
Plugging (1.22) into (1.20), they obtained
\be
[X^{\mu}(\sigma,\tau),X^{\nu}(\sigma',\tau)]=\left\{\bea{cc}
-2\pi i\alpha'({\cal M}^{-1}{\cal F})^{\mu\nu} ,\;{\rm for}\;\;  \sigma=\sigma'=0 \\ \;
2\pi i\alpha'({\cal M}^{-1}{\cal F})^{\mu\nu} ,\;{\rm for}\;\; \sigma=\sigma'=\pi \\ \;
0\;\;\;\;\; otherwise.
\eea\right.
\ee
In other words, they concluded that mixed open strings are noncommutating only
at the ends, where open strings are attached to the brane, i.e.  
$\sigma=\sigma'=0\; ,\; \sigma=\sigma'=\pi$, and the center of mass 
coordinates are commuting. 

To settle the question of the seat of noncommutativity in a more fundamental 
point of view, since the problem arises from the boundary conditions which can 
be treated
as constraints, we take the Dirac's quantization method to analyze 
the problem [17], which we will now outline:

Suppose we take a system with the canonical position and momenta 
$\; (q_i,p_i);\; i=1,...,n$ and assume that they are constrained to lie on the hypersurfaces 
$\Phi_m(q,p)=0\; , m=1,...,r$. In studying this system we face two 
problems; 

i) The Hamiltonian is not well-defined and can be modified by adding 
a linear combination of the constraints; and

ii) The dynamics may not respect the constraints. 

In the Dirac's method both difficulties are resolved. First, if the  
constraints are violated under time evolution, by taking the Poisson bracket
of Hamiltonian and the constraints, we may construct a larger set of  
constraints which are consistent with dynamics. Furthermore the freedom in the
Hamiltonian may also be fixed by this consistency. The final step to construct the
Dirac brackets is to check the consistency of the constraints, i.e.  consider 
the Poisson brackets of the constraints. If they turn out to be zero, the 
constraints are called first class and they are consequence of symmetries of 
the system. If it happens that the constraints have non-vanishing Poisson brackets, 
and the matrix of their brackets is non-singular, the second class
constraints, Dirac defines the quantum commutator in terms of a modified 
Poisson bracket, i.e. Dirac bracket. 
It is interesting that 
in the case of second class \cons we can find a new set of canonical variables 
in which the \cons are practically decoupled from the physical degrees of 
freedom [23,24] and the Poisson brackets of the new set of variables are
the same as
Dirac brackets. We take the \BCs as  \cons and follow the procedure  
described above. 
The \cons are of second class. We find that the coordinates at the ends points 
of the strings are \nc. The exact value of the commutators depend on the 
$\Lambda$-gauge we choose, but as we can see from its expression given below,
is non-zero in any gauge
\footnote{ we will present the calculations in section 4.};
\be
[X^i(0,\tau),X^j(0,\tau)]={2\pi\alpha'i}\biggl((-{\cal M}+{1\over 2}
{F}{\cal F})^{-1}{\cal F}\biggr)^{ij}.
\ee
This is unlike the commutators of the momenta which can be set to zero for
all string points in the gauge $F=0$. This gauge is particularly useful 
since the mixed open strings momentum distribution does not possess delta 
function singularity at the end of the string. 

In section 2, we review the method of Dirac quantization in detail. In
section 3, we  apply it to the well-known cases of D-branes and show the
equivalence of the Dirac method and the usual mode expansion and
quantizing the
Fourier modes. In section 4, applying 
Dirac's method to the case of non-zero background antisymmetric field, i.e. 
the case of mixed boundary conditions, we show that despite gauge 
dependence, the noncommutativity is not removable. 
In the Dirac's prescription, unlike the mode expansion method, 
the boundary conditions at two ends of open strings appear independently, so
we can consider the strings with their ends on the same or different branes.  
In this section , we also compare our results with the previous
methods, and argue that in a gauge in which the internal $U(1)$ field is
set to zero, the results of Dirac quantization agree with the mode expansion
method provided we set the commutators of the center of mass coordinates 
to zero.
The last section is devoted to conclusions and discussions.

\section{Review of Dirac Quantization}
\setcounter{equation}{0} 
  According to principles of quantum mechanics, to quantize a mechanical
system 
a Hamiltonian formulation of the system is needed, i.e. a Hamiltonian 
and a Poisson bracket structure. Having these two we can find the dynamics of 
any function defined on the phase space of the theory, and to go over to the  
quantized theory we should replace the Poisson bracket with the commutators. But,
in general, when the theory is subject to constraints, i.e.
when all of the phase space is not available for the motion, the 
Hamiltonian formulation should be modified. Dirac has given a concise recipe 
to handle these constrained systems [17], which we briefly review here. 

Given a Hamiltonian, $H$, and a set of {\it primary} constraints, 
$\Phi_m(q,p)=0$, since any linear combination of the constraints is also zero, 
the Hamiltonian is not uniquely determined and $H_T=H+u_m\Phi_m$
is as good as $H$, 
$u_m$ being arbitrary functions in phase space. 
Dynamics of any function in phase space, $g(q,p)$, is determined by
the Poisson bracket structure, 
\be
{\dot g}\approx \{ g, H_T\}_{P.B.},
\ee
the symbol $\approx$ called weak equivalence implying that, 
the Poisson brackets should be worked out  and then the constraints imposed, 
thus
\be
\Phi_m\approx 0.
\ee
Consistent dynamics requires that the constraints remain valid under time 
evolution:
\be
{\dot \Phi_m}\approx \{ \Phi_m, H_T\}_{P.B.}\approx\{\Phi_m,H \}+u_n
\{\Phi_m,\Phi_n\}\approx 0.
\ee

There are then three possibilities 

{\it i)} The equation (2.3) reduces to $0=0$, i.e. identically 
satisfied with the imposition of the primary constraints.

{\it ii)} They reduce to equations independent of $u$'s, thus involving only 
q's and p's. Such equations must be independent of the primary constraints,
otherwise they would be of the first kind, thus they are  a new set of 
constraints:
\be
\Psi_i(q,p)=0.
\ee

{\it iii)} The equations (2.3) may not reduce in either of these ways; they then
may be solved in terms of the $u$'s.

Equations obtained from the kind {\it ii)}, (2.4), 
are called {\it secondary} constraints. If 
a secondary constraint turns up, then their 
consistency with the time evolution should still be worked out:
\be
{\dot \Psi_i}\approx\{ \Psi_i, H_T\}_{P.B.}\approx 0.
\ee
These equations should be treated on the same footing as (2.3), and the
analysis should be repeated. The procedure should be
pursued until there are no more secondary constraints and we have
exhausted all the consistency conditions. Finally we will be left with a
number of secondary constraints and a number of conditions on $u$'s.
The secondary constraints, for many purposes should be treated on the same 
footing as the 
primary ones; e.g. , in evaluating the weak equations they should be
imposed 
after 
calculating the Poisson brackets. So for an arbitrary constraint,
$\chi(q,p)$,
primary or secondary:
\be
\{ \chi_i, H_T\}_{P.B.}\approx 0.
\ee

There is a more important classification of constraints: 
\newline
A constraint is called {\it first class}, if its Poisson bracket with all other  
constraints vanishes, a well-known example of which is the electromagnetic 
field [23].
First class constraints arise from symmetries of the action which in the 
electromagnetic case is the gauge invariance.
\newline
The {\it second class} constraints are those with a non-zero and 
non-singular Poisson brackets, 
\be
\{ \chi_M,\chi_N\}_{P.B.}\equiv C_{MN},\;\;\; 
\ee
with $det\;C\neq 0$.

According to Dirac, to quantize a system with second class constraints,
the commutators of any two functions on the phase space should be 
given in terms of a modified bracket, Dirac bracket,
\be
[A,B]=i\{A,B\}_{D.B.},
\ee
where 
\be
\{A,B\}_{D.B.}=\{A,B\}_{P.B.}-\{A,\chi_M\}_{P.B.}\;(C^{-1})^{MN}
\{\chi_N,B\}_{P.B.}.
\ee
One can easily verify that the Dirac bracket satisfies the algebraic
relations of the Poisson bracket.
The important property of the Dirac bracket is that for an arbitrary $A$ and
for all constraints $\chi_M$
\be
\{\chi,A\}_{D.B.}=0.
\ee
So, using the Dirac brackets instead of Poisson bracket, the weak equations 
may be written as strong equalities.

For second class constraints we can always
find a {\it canonical transformation} which maps the set of the constrained
variables, $q_{\alpha},p_{\alpha}\; ,\alpha=1,...,N,$ and the constraints 
$\chi_M,\; ,M=1,...,n$, to 
a set $Q_{i},P_{i};\xi_M\; ,i=1,..,N-n, $ , such that $\xi_M=0$ identically, 
and $Q,P$ are unconstrained [24], i.e.
\be
\{A,B\}_{D.B.}=\{A,B\}_{P.B. {\rm in\; terms\; of}\; Q,P}
={\partial A\over \partial Q_i}{\partial B\over \partial P_i}-
{\partial B\over \partial Q_i}{\partial A\over \partial P_i}.
\ee
The set of $(Q,P)$ span the {\it reduced phase space}, and it has been shown
that the quantization on this reduced phase space with the canonical commutators 
is equivalent to the Dirac quantization on the constrained phase space.

\section{Boundary Conditions as Dirac Constraints}
\setcounter{equation}{0} 
 In this section, having equipped ourselves with the Dirac bracket, we return 
to the problem of quantization of usual open strings which in the simplest case 
is governed by 
\be
S= {1 \over 4\pi\alpha'} \int \bigl( 
(\partial_{\tau}X^{i})^2-(\partial_{\sigma}X^{i})^{2}\bigr) d\sigma d\tau. 
\ee
Variation of the above action reduces to the equations of motion: 
$\partial_{\tau}^2X^{i}=\partial_{\sigma}^2X^{i}$ and 
the Neumann boundary conditions, $\partial_{\sigma}X^{i}|_{0,\pi}=0$. Treating
the boundary conditions as the Dirac constraints, we try to
quantize the theory directly by means of the Dirac method, without recourse to 
the mode expansions. Our point of view here, treating the boundary 
conditions as constraints, is justified and discussed in more detail in [25].

The total Hamiltonian of the open strings is
\be
H= {1 \over 4\pi\alpha'} \int \bigl( (\partial_{\tau}X^{i})^2
+(\partial_{\sigma}X^{i})^{2}\bigr) d\sigma d\tau+u_i
\partial_{\sigma}X^{i}|_{0\; or\pi},
\ee
with the canonical momenta
\be
2\pi\alpha'P^i(\sigma,\tau)=\partial_{\tau}X^{i}.
\ee
and the primary constraints: 
$\;\;\Phi_i\equiv\partial_{\sigma}X^{i}|_{0}=0$. 
The boundary conditions at the other end of the string $(\sigma=\pi)$ can be worked 
out similarly.

Now, the consistency of the constraints,
\be
\{ \Phi_i, H\}_{P.B.}=2\pi\alpha'\partial_{\sigma} P^i|_{0},
\ee
gives a new set of constraints; and therefore it is  of the {\it ii)} kind. 
Calling them $\Psi_i$, we can go on
$$
\{ \Psi_i, H\}_{P.B.}\approx 0,
$$
and find an equation for $u_i$. As discussed in [25], although $u_i$ is 
determined the constraint chain is not terminated, but along the arguments of
[25] taking only $(\Phi_i, \Psi_i)$ from the constraint chain will give the 
proper result if at the end, we reconsider a normalization factor for $X_i$ and 
$P_j$ so that, their Dirac brackets (or commutators) become $\delta_{ij}$.
Since the Poisson brackets of the constraints are not zero,
we are dealing  with second class constraints,
\be\bea{cc}
\{ \Phi_i,\Phi_j\}_{P.B.}=0\;\;\; ,\;\;\; \{ \Psi_i,\Psi_j\}_{P.B.}=0\\
\{\Phi_i,\Psi_j\}_{P.B.}=\{\partial_{\sigma}X^i|_0,\partial_{\sigma}P^j|_0\}_{P.B.}
=\int_0^{\pi}\{\partial_{\sigma} X^i(\sigma),\partial_{\sigma}P^i(\sigma')\}
_{P.B.}\delta(\sigma)\delta(\sigma')d\sigma d\sigma'.
\eea\ee
To overcome the difficulty of dealing with the delta function derivatives at 
the boundaries, we discretize $\sigma$, denoting the
steps by $\epsilon$, and take the $\epsilon \rightarrow 0$ 
at the end; thus
\be
\Phi^i={1\over \epsilon}(X^i_1-X^i_0).
\ee
With the discrete Lagrangian conjugate momenta:
\be
2\pi\alpha' P_0^i=\epsilon {\dot X_0^i}\;\;,\;\;
2\pi\alpha' P_1^i=\epsilon {\dot X_1^i}.
\ee
The $X_n^i,P_n^i,\; n=0,1$ have the usual Poisson brackets
\be\bea{cc}
\{X_n^{i},X_m^{j}\}=0, \\ \;\;\;
\{P_n^{i},P_m^{j}\}=0 \\   \;\;\;
\{X_n^{i},P_m^{j}\}=\delta^{ij}\delta_{mn}.
\eea\ee
The secondary constraints in the discretized version are
\be
\Psi^i=P_1^{i}-P_0^{i},
\ee
hence
\be
\{\Phi_i,\Psi_j\}_{P.B.}={1\over \epsilon}\{X^i_1-X^i_0, P_1^{j}-P_0^{j}\}=
{2\over \epsilon}\delta^{ij}. 
\ee
It should be emphasized that, we use the discrete language only as a method for  
calculating the commutators and at the end the objects surviving in the 
$\epsilon \rightarrow 0$ limit are physical.
\newline
The $C$ matrix takes the form 
\be
\left(\bea{cc} \; 0\;\;{2\over \epsilon}\delta^{ij}\\
{-2\over \epsilon}\delta^{ij}\;\;\;\ 0
\eea\right),
\ee
and is invertible. For simplicity and to visualize how the Dirac method goes
through, let us take the boundary conditions only on one of $X^i$'s.
In this case $C$ is a $2\times 2$ matrix and
\be
C^{-1}={\epsilon\over 2}\left(\bea{cc} \; 0\;\;\;-1\\  1\;\;\;\; 0
\eea\right).
\ee
To quantize the theory we calculate the Dirac brackets and find
\be
\{X^i(\sigma),X^j(\sigma')\}_{D.B.}=0-\{X^i(\sigma),\chi_M\}_{P.B.}\;(C^{-1})^{MN}
\{\chi_N,X^i(\sigma)\}_{P.B.} =0.
\ee
The Dirac bracket for two $P^i(\sigma)$'s similarly vanishes, but 
$\{X^i(\sigma),P^j(\sigma')\}_{D.B.}$ will have a slightly different form near the
boundary, i.e. at $\sigma,\sigma'=0,\epsilon$ which can be removed by
normalizing $X$ and $P$ by a factor of $\sqrt 2$. The above Dirac brackets
can be understood in terms of Maskawa-Nakajima theorem, [24] too.
We can easily find the 
reduced phase space and the unconstrained variables, which are 
$$
X^i={X_1^{i}+X_0^{i} \over 2}\;\;\;, \;\;\; P^i=P_1^{i}+P_0^{i},
$$
and the identically vanishing constraints: 
$$
X_1^{i}-X_0^{i} \;\;\;, \;\;\; P_1^{i}-P_0^{i}.
$$

Comparing the method we used here with the usual open strings quantization based 
on the mode expansion, we find that in the mode expansion method, to impose 
the boundary conditions, we put coefficients of the Sine terms in the 
expansion equal to zero. 
This is equivalent to considering the secondary constraints and {\it also} 
going to {\it the reduced phase space}, and performing the quantization on these 
unconstrained set of variables. Hence in the Neumann boundary conditions, 
the mode expansion quantization and the Dirac method give the same quantum theory.

One can similarly apply the Dirac bracket method to the strings with  
Dirichlet boundary conditions. Again there are secondary constraints
and the $C$ matrix is invertible, giving 
$\{X^i(\sigma),X^j(\sigma')\}_{D.B.}=0$,
$\{P^i(\sigma),P^j(\sigma')\}_{D.B.}=0$.
The reduced phase space is the one on which the $X_0, P_0$ are both omitted,
which is again the same as what we have in the mode expansions; there we only
choose the Sine, which is equivalent to working in the reduced phase space.

\section{Dirac Quantization For Mixed Boundary Conditions}
\setcounter{equation}{0} 
 In this section we consider a more general case, the open strings with 
mixed boundary conditions.
Let us take the open strings ending on a p-brane which is spanned by 
$X^0,X^1,...,X^p$, assuming that we have turned on a constant $B_{\mu\nu}$ 
$(\mu,\nu=0,...,p)$ background and also we have a constant $F_{\mu\nu}$ 
ending on the brane. As
explained earlier since the background $B$ is constant, it does not appear in
the equation of motion and only appears in the boundary conditions
\be\bea{cc}
\;\;\; \partial^2_{\tau}X^{\nu}=\partial^2_{\sigma}X^{\nu}, \\
\partial_{\sigma}X^{\mu}+{\cal F}_{\mu\nu} \partial_{\tau}X^{\nu}=0 \;\;\;
{\rm at}\;\; \sigma=0,\pi.
\eea\ee
The above equations enjoy two $U(1)$ symmetries explained earlier and hence
the solution of them also have these symmetries. Calculating the conjugate
momenta of $X^{\mu}$
\be
2\pi\alpha' P^{\mu}(\sigma,\tau)=\partial_{\tau}X^{\mu}+
B^{\mu}_{\nu}\partial_{\sigma}X^{\nu}+A^{\mu}(\delta(\sigma-\pi)-\delta(\sigma)),
\ee
we see that the momentum is not gauge invariant under either of these
$U(1)$'s. The $U(1)$ acting only on $A$, is similar to the familiar 
electrodynamics gauge symmetry and at the end survives in our quantized
theory. 
Hence by choosing a gauge, we fix the $U(1)$ which 
only acts on $A^{\mu}$ field (1.8). For constant $F_{\mu\nu}$, $\lambda$ 
can be chosen so that
\be
A_{\mu}={-1\over2}F_{\mu\nu}X^{\nu}.
\ee
To quantize the theory, we consider the mixed 
boundary conditions as constraints and quantize the system by the Dirac method.
Running the Dirac machinery first we should build the Hamiltonian from the 
action:
\be
H= {1 \over 4\pi\alpha'} \int \bigl( (\partial_{\tau}X^{\mu})^2
+(\partial_{\sigma}X^{\mu})^{2}\bigr) d\sigma.
\ee
$\partial_{\tau}X^{\mu}$ are related to momenta through (4.2). We will  
put $2\pi\alpha'$ equal to one and reintroduce in our results at the end.
Without loss of generality, we assume the electric mixing, 
i.e. ${\cal F}_{0\nu},{B}_{0\nu}$ to be zero. The boundary conditions,
or in other words, the primary constraints are 
\be
\Phi^i=\partial_{\sigma}X^{i}+{\cal F}_{ij} \partial_{\tau}X^{j}|_
{\sigma=0\;\;\ or\;\pi}=0. \;\;\;\; i,j=1,2,...,p.
\ee
Here the calculations are presented explicitly only for $\sigma=0$; for the 
other end ($\sigma=\pi$), which is very similar to $\sigma=0$, we only give the
results.

According to Dirac prescription, we should verify if there are any other 
constraints:
\be
\{ \Phi_i, H_T\}_{P.B.}=\int_0^{\pi}\{\partial_{\sigma}X^{i}(\sigma)
+{\cal F}_{ij} \partial_{\tau}X^{j}(\sigma),H\}\delta(\sigma)d\sigma+ 
u_j\{ \Phi_i,\Phi_j\}_{P.B.}.
\ee
The $\{ \Phi_i,\Phi_j\}_{P.B.}$ in the above is ill-defined; hence the
two terms should vanish independently; which leads to secondary constraints [25].

To avoid the ambiguities due to $\delta(\sigma)$ derivatives, 
we use the discretized form of the strings:
\be
\Phi^i= X_1^{i}-X_0^{i} +\epsilon{\cal F}_{ij} \partial_{\tau}X^{j}_0,
\ee
and the related part of the Hamiltonian
\be
H={\epsilon\over 2}({\dot X_0^2}+{\dot X_1^2})+{1\over 2\epsilon}
[(X_1-X_0)^2+(X_2-X_1)^2],
\ee
with
\be\bea{cc}
P^i_0=\epsilon{\dot X_0^i}+ B_{ij}(X_1-X_0)_j+{1\over2} F_{ij}X_{0j}\\
P^i_1=\epsilon{\dot X_1^i}+ B_{ij}(X_2-X_1)_j.
\eea\ee
$P^i_n,X^i_n ,\; n=0,1\;$ have the usual canonical Poisson brackets (3.7).

The secondary constraints are
\be
\Psi^i=\{ \Phi_i, H\}={1 \over \epsilon}({\bf 1}- {\cal F}B)^{ij}
(P_1^{j}-P_0^{j}+{1\over2} F_{jk}X_{0k})=0,
\ee
which are not invariant under $\Lambda$-gauge transformation (1.9) explicitly.
We must also check if there is any more constraints; this leads to an infinite
chain of constraints, but as discussed in [25] taking all these constraints
into account is equivalent to take only $(\Phi_i, \Psi_i)$ and normalize $X$
and $P$ with the same numeric factor so that $[X^i_n,P^j_m]=1\delta_{mn}
\delta^{ij}$.

At this stage following Dirac, the class of the constraints
should be determined;
\be\bea{cc}
A_{ij}\equiv\{ \Phi_i,\Phi_j\}_{P.B.}=2{\cal F}^{ik}{\cal M}^{kj}-
{\cal F}^{ik}{F}^{kl}{\cal F}^{lj},\\
D_{ij}\equiv\{ \Phi_i,\Psi_j\}_{P.B.}=2{\cal M}^{ij}-
{\cal F}^{ik}{F}^{kj}, \\
\{ \Psi_i,\Psi_j \}=-F_{ij}
\eea\ee
with
\be
{\cal M}^{ij}=({\bf 1}- {\cal F}^2)^{ij}.
\ee
Since the Poisson brackets of constraints are non-zero we may be dealing with 
{\it second class} constraints; to check that and to quantize the system we 
should find the $C$ matrix, the Poisson brackets of the constraints.
In the general p-brane case $C$ is a $2p\times 2p$ matrix composed of $A,D,F$ 
matrices defined in (4.11):
\be
C=\left(\bea{cc} A\;\;\;\ D\\
-D^t\;\;\ -F
\eea\right).
\ee
\be
det C=det (-AF+DD^t).
\ee
The above determinant, like the $C$ itself, is $\Lambda$-gauge dependent, 
but in any gauge, for real $\Lambda$, (4.14) is non-zero and hence, $C$
is invertible. Therefore the constraints are in fact 
second class. To visualize how the Dirac method works, without loss of 
generality, we confine ourselves to the $D_2$-brane, i.e. $p=2$, in which
$C$ is a $4\times 4$ matrix.  
We are now ready to calculate the basic Dirac brackets:
\be
\{X^i_0,X^j_0\}_{D.B.}=\biggl((-2{\cal M}+
{\cal F}{F})^{-1}{\cal F}\biggr)^{ij},
\ee
\be
\{X^i_0,X^j_1\}_{D.B.}=0,
\ee
\be
\{X^i_1,X^j_1\}_{D.B.}=\biggl((2{\cal M}-
2{\cal F}{F})^{-1}{\cal F}\biggr)^{ij}.
\ee
Dirac brackets of $X^i$ for any point in the middle of the string are zero.

The important and new result we obtain here is that the Dirac brackets are 
{\it not} gauge invariant, and each of (4.15) or
(4.16) could be set to zero by a gauge transformation. But, there is no gauge 
in which both of these brackets vanish. 

Similarly we can find the basic Dirac brackets containing the momenta:
\be\bea{cc}
\{P^i_0,P^j_0\}_{D.B.}= {1\over 4}F^{ij}, \\ \;
\{P^i_0,P^j_1\}_{D.B.}=0, \\ \;
\{P^i_1,P^j_1\}_{D.B.}={-1\over 2}
\bigl(({\cal M}-{\cal F}F)\;F\;(2{\cal M}-F{\cal F}))^{-1}\bigr)^{ij}.
\eea\ee
and
\be\bea{cc}
\{X^i_0,P^j_0\}_{D.B.}={1\over 2}\delta^{ij} \;\;\;\; ,\;\;\;\;
\{X^i_0,P^j_1\}_{D.B.}={1\over 2} \\ \;
\{X^i_1,P^j_0\}_{D.B.}={1\over 2}\delta^{ij} \;\;\;\; ,\;\;\;\; 
\{X^i_1,P^j_1\}_{D.B.}={1\over 2}\delta^{ij}.
\eea\ee

To find the proper Dirac brackets, we should normalize the $X^i_n$ and 
$P^j_m$ by a factor of $\sqrt 2$ so that the brackets of (4.19) are in their
usual form. Hence the brackets of (4.17) and (4.18) should be multiplied by a 
factor of 2.

The {\it quantum theory} is obtained by replacing the Dirac brackets with 
-i$\times$commutators. Reintroducing the $2\pi\alpha'$ factor we have

\be\bea{cc}
[X^i_0,X^j_0]={-2\pi\alpha' i}({\cal M}-{1\over 2}{\cal F}{F})^{-1}{\cal F}, \\ \;
[X^i_0,X^j_1]=0, \\ \; 
[X^i_1,X^j_1]={2\pi\alpha' i}({\cal M}-{\cal F}{F})^{-1}{\cal F}.
\eea\ee

In other words the open strings coordinates on the brane, along which 
$B_{ij}$ and $F_{ij}$
have a non-zero value, become {\it noncommuting}, and the noncommutativity 
being {\it gauge dependent}, but not removable.

A special gauge which has been discussed earlier in [8,9] is:
$$
F=0 \;\;\; or \;\;\; B={\cal F}.
$$
In this gauge commutators take a simpler form
\be\bea{cc}
[X^i_0,X^j_0]=-[X^i_1,X^j_1]={-2\pi\alpha' i}({\cal F}{\cal M}^{-1})^{ij},\\ \;
[X^i_0,X^j_1]=0,
\eea\ee
and
\be
\bea{cc}
[X^i_n,P^j_m]=\delta^{ij}(\delta_{n0}+\delta_{n1})(\delta_{m0}+
\delta_{m1}), \;\; \\ \;
[P^i_n,P^j_m]=0.
\eea\ee
In the last part of this section we will compare our results in this gauge 
($F=0$) with the previous works, based on the mode expansions.
\newline
{\it Boundary Conditions at $\sigma=\pi$}

To discuss the noncommutativity of open strings center of mass we should
also consider the boundary conditions at the other end of the open string. 
There are some choices: The boundary conditions can be Neumann, Dirichlet or 
mixed with the same or different mixing parameter, ${\cal F}'_{ij}$. The open 
strings with Mixed boundary conditions at one end ($\sigma=0$), and Neumann 
(Dirichlet) at the other end ($\sigma=\pi$) are reffered to as $MN$ ($MD$) 
strings, and the open strings
with mixed boundary conditions at both ends but with different (or the same)
mixing parameters, $MM'$ ($MM$).
The first two cases, $MN$ and $MD$, happen when a mixed p-brane
is parallel to a $D_p$-brane and $D_{p-2}$-brane respectively [5].
$MM$ strings appear when the open string has both ends on the same
mixed brane.

To study the most general case we consider the $MM'$ strings which
recovers the $MM$, $MD$ and $MN$ as special cases. Since the calculations
are very similar to what we have done earlier, we will only quote the 
results here
\be
\{X^i_{\pi},X^j_{\pi}\}_{D.B.}=
\biggl((2{\cal M}'-{\cal F}'{F'})^{-1}{\cal F}'\biggr)^{ij},
\ee
\be
\{X^i_{\pi},X^j_{\pi-\epsilon}\}_{D.B.}=0,
\ee
\be
\{X^i_{\pi-\epsilon},X^j_{\pi-\epsilon}\}_{D.B.}=
\biggl((-2{\cal M}'+2{\cal F}'{F'})^{-1}{\cal F}'\biggr)^{ij},
\ee
Equations (4.23-25) compared to similar results at $\sigma=0$, (4.15-17), have 
an extra minus sign. This can be understood intuitively noting that 
the open strings attached to branes are oriented open strings. 
\newline
{\it Noncommutativity of Open string Center of Mass}

For $MM'$ strings discussed above the commutator of the center of mass
coordinates, 
$$
X^i_{c.m.}=\int_0^{\pi} d\sigma\;\; X^i(\sigma,\tau)=\epsilon \sum_{m=0}^N
X^i_m,
$$
is found to be
\be
[X^i_{c.m.},X^j_{c.m.}]={\epsilon}^2 \sum_{m,n=0}^N [X^i_n,X^j_m]=
{4\pi\alpha'i}\bigl\{{1\over \sqrt{det C}}({\cal F}{F}{\cal F})^{ij}-
{1\over \sqrt{det C'}}({\cal F}'{F'}{\cal F}')^{ij}\bigr\},
\ee
where $det\; C'$ has the same form of $det\; C$ (4.14), with ${\cal F}$ replaced by   
${\cal F}'$. The above relation is in general {\it $\Lambda$-gauge dependent}. 
But, in the important special case when the open string has both ends on the
same mixed brane ($MM$ string), the \ncy at the ends of these oriented
open
strings cancel each other and the center of mass commutators therefore 
{\it vanishes} and {\it is not gauge dependent}. In the other special cases, 
$MN$ and $MD$ open
strings, we can choose a gauge ($F=0$) in which the center of mass coordinates
are commuting [5].

The \ncy of $MM$ open string components was first observed in [7], and in a 
more detailed work discussed in [8], and later in [9]. 
In the introduction we noted that the quantization methods
used in [8] and [9], both based on the string mode expansions, 
and with the choice (1.22) are exactly equal.
In this part we compare the results of the Dirac quantization method, with the 
previous works. 

In order to make this comparison the Dirac method results should be rewritten in the 
special $\Lambda$-gauge, $F=0$, which leads to (4.21), (4,22) and also 
be translated into the continuum language:
\be\bea{cc}
X^i_0=X^i(\sigma,\tau)|_{\sigma=0} \;\;\;\;\; , \;\;\;\;\; X^i_1=X^i_0+
\epsilon\partial_{\sigma}X^i|_{\sigma=0} \\ 
P^i_0=\epsilon P^i(\sigma,\tau)|_{\sigma=0} \;\;\;\;\; , \;\;\;\;\; P^i_1=P^i_0+
{\epsilon}^2\partial_{\sigma}P^i|_{\sigma=0},
\eea\ee
and the similar relations at $\sigma=\pi$.
Inserting these into (4.21,23) we obtain
\be\bea{cc}
[X^i(\sigma,\tau)|_{\sigma=0},X^j(\sigma,\tau)|_{\sigma=0}]=-
[X^i(\sigma,\tau)|_{\sigma=\pi},X^j(\sigma,\tau)|_{\sigma=\pi}]
={-2i\pi\alpha'}({\cal F}{\cal M}^{-1})^{ij},\\ \;
[\partial_{\sigma}X^i|_{\sigma=0},\partial_{\sigma}X^j|_{\sigma=0}]=0,
\eea\ee
and for the $X,P$ commutators,
\be
\bea{cc}
[X^i|_{\sigma=0},P^j|_{\sigma=0}]={1\over\epsilon}\delta^{ij}, \;\; \\ \;
[\partial_{\sigma}X^i|_{\sigma=0}, P^j|_{\sigma=0}]=0,\;\;\;  \\ \;
[\partial_{\sigma}X^i|_{\sigma=0}, \partial_{\sigma} P^j|_{\sigma=0}]=0,
\eea\ee
and also
\be
[P^i|_{\sigma=0},P^j|_{\sigma'=0}]=0.
\ee
$[X^i|_{\sigma=0},\partial_{\sigma}X^j|_{\sigma=0}]$ in the 
$\epsilon\rightarrow 0$ (continuum limit) is not finite, and 
tends to infinity as ${1\over \epsilon}$ .

Comparing the Dirac's method results in the $\Lambda$-gauge in which $F$ is 
zero, (4.28), with (1.23) one finds that they are in exact agreement.

The other commutators, i.e. 
$[\partial_{\sigma}X,\partial_{\sigma}X],\;[\partial_{\sigma}X,P],\;
[\partial_{\sigma}X,\partial_{\sigma}P],\;[P,P]$, can be worked out in the
mode expansion method. Doing so, which is a straightforward calculation, we 
find the same results in both methods, the Dirac's and the mode expansion.
One can go further and find all the commutators of mode expansion
quantization from the Dirac method by considering the general $\sigma$, not only 
$0$ and $\pi$, which we will not do it here.
\newline
Thus we see that the quantization methods based on mode expansion, 
[8] and [9], which rely on certain ad-hoc assumptions agree with the more 
fundamental one, the Dirac quantization procedure.
\section{Conclusions and Discussions}
\setcounter{equation}{0} 
We have tackled the problem of the \ncy of the internal coordinates of brane 
starting from the basic principles of  Dirac quantization. We found that 
in spite of a large class of gauge symmetries, and gauge dependence of the
form  of the \ncy of the open strings attached to the brane 
in the presence of constant background antisymmetric field, there is no gauge
in which the \ncy can be removed, i.e., \ncy of the coordinates of the open 
string-brane system is {\it intrinsic}. As the gauge changes, the details
of the \ncy is altered, but it never goes away altogether.
The position of noncommutativity is always on the brane. 
We discussed that, choosing the mixed \BC at one end, there 
are various choices for the \BC at the other end of open strings; 
it can be Neumann, Dirichlet,
and Mixed with the same or different mixing parameters, referred to as $MN$, 
$MD$, $MM$ and $MM'$ respectively. And showed that except for the $MM$ case,
the center of mass coordinates of these open strings are noncommuting but
$\Lambda$-gauge dependent, so that the center of mass \ncy can be
removed in a certain gauge. 

The $MM$ case physically describes the open strings having both ends on a 
mixed brane. For this case we showed that, although the open string
end points are not commuting and this \ncy is $\Lambda$-gauge dependent,
the center of mass coordinates are commuting which is a gauge independent
result.

Having obtained the Connes' \nc structure as an inherent property of
certain branes, many open questions can be addressed.
The \ncy arguments based on mixed open strings quantization hold in both 
compact and non-compact branes. In the compact case [8], we showed that the 
$C^*$-algebra of the functions on \nct can be constructed through the open 
string zero modes. However, in the non-compact case, in spite of having \nc
coordinates, since these functions are not normalizable, there is no 
$C^*$-algebra structure. In this case the brane is a {\it Moyal plane}.
It is shown [14,15] that the internal low energy effective theory living in
this Moyal plane is \nc super Yang-Mills theory described by Moyal brackets 
[10]. Renormalizability of this theory, which has been checked up to one loop,
is a question to be answered.

The other problem to be addressed is the dynamics of branes in \nc backgrounds.
For the special case, zero-branes on a \nc two-torus, corresponding dynamics 
has been proposed to be   
a light cone formulation of M-theory with a non-zero three form; a generalization
of Matrix-model conjecture. But, there are some suggestions that the Born-Infeld
corrections to the Yang-Mills are not suppressed in the large $N$ limit, 
unlike the commutative case. Hence to have a proper formulation of Matrix-theory, 
Born-Infeld action on the \nc background should be considered and all the 
Matrix-theory arguments should be repeated for the \nc Born-Infeld,
some of which have been considered by Hofman-Verlinde.

Another interesting open problem is the anomalous $U(1)$ symmetry. 
${\cal F}=B-dA$, being invariant under both $\lambda$ and $\Lambda$ gauge 
symmetries, is a good parameter to describe the system at the classical level, 
but according to our explicit calculations 
these gauge symmetries do not survive at the quantum level. This can be 
understood noting that at the quantum level, the $\lambda$ gauge symmetry 
is lifted to the deformed U(1), which is the isometry group of the Moyal plane 
[20]. It seems that the supersymmetry arguments can 
help to understand this anomaly. Resolving this problem may shed light on the
problem of $U(N)$ Born-Infeld theory and also the noncommutative Yang-Mills or DBI 
action, which is expected to have $\Lambda$-symmetry.

\vskip 2cm

{\bf Acknowledgements}
\newline
M.M Sh-J. would like to thank A. Shirzad for many helpful comments and 
discussions.

\end{document}